\documentclass[Crown,doublespace,sagev,times]{sagej}
\usepackage{array}
\usepackage{multirow}
\usepackage{amsmath}
\usepackage{amssymb}
\usepackage[colorlinks,bookmarksopen,bookmarksnumbered,citecolor=red,urlcolor=red]{hyperref}
\usepackage{url}
\usepackage{adjustbox}
\usepackage{caption}
\usepackage{subcaption}

\newcommand\BibTeX{{\rmfamily B\kern-.05em \textsc{i\kern-.025em b}\kern-.08em
T\kern-.1667em\lower.7ex\hbox{E}\kern-.125emX}}

\def\volumeyear{2020}

\begin{document}

\runninghead{E.R. van den Heuvel et al.}

\title{Simulation Models for Aggregated Data Meta-Analysis: Evaluation of
Pooling Effect Sizes and Publication Biases.}

\author{Edwin R. van den Heuvel\affilnum{1,2}, Osama Almalik\affilnum{1},
and Zhuozhao Zhan\affilnum{1}}

\affiliation{\affilnum{1}Department of Mathematics and Computer Science, Eindhoven University of Technology, Eindhoven, The Netherlands\\
\affilnum{2}Department of Preventive Medicine and Epidemiology, School of Medicine, Boston University, Boston, USA}

\corrauth{E. R. van den Heuvel, Department of Mathematics and Computer Science, Eindhoven University of Technology, Den Dolech 2, 5612 AZ, Eindhoven, The Netherlands.}

\email{e.r.v.d.heuvel@tue.nl}

\begin{abstract}
Simulation studies are commonly used to evaluate the performance
of newly developed meta-analysis methods. For methodology that is
developed for an aggregated data meta-analysis, researchers often
resort to simulation of the aggregated data directly, instead of simulating
individual participant data from which the aggregated data would be
calculated in reality. Clearly, distributional characteristics of
the aggregated data statistics may be derived from distributional
assumptions of the underlying individual data, but they are often
not made explicit in publications. This paper provides the distribution
of the aggregated data statistics that were derived from a heteroscedastic
mixed effects model for continuous individual data. As a result, we
provide a procedure for directly simulating the aggregated data statistics.
We also compare our distributional findings with other simulation
approaches of aggregated data used in literature by describing their
theoretical differences and by conducting a simulation study for three
meta-analysis methods: DerSimonian and Laird's pooled estimate and
the Trim \& Fill and PET-PEESE method for adjustment of publication
bias. We demonstrate that the choices of simulation model for aggregated
data may have a relevant impact on (the conclusions of) the performance
of the meta-analysis method. We recommend the use of multiple aggregated
data simulation models for investigation of new methodology to determine
sensitivity or otherwise make the individual participant data model
explicit that would lead to the distributional choices of the aggregated
data statistics used in the simulation.
\end{abstract}


\maketitle

\section{Introduction}

An aggregated data meta-analysis would collect the information $(D_{i},S_{i},df_{i})$
for study $i,$ with $D_{i}$ the estimated study effect size of interest
(e.g., mean difference, Cohen's $d$, log odds ratio), $S_{i}$ the
estimated standard error of the study effect size, and $df_{i}$ the
corresponding degrees of freedom for the standard error\footnote{Depending on the meta-analysis approaches or applications, the degrees
of freedom $df_{i}$ may or may not play a role in the analysis. For
instance, the degrees of freedom is used in an investigation of residual
heteroscedasticity with Bartlett's test when studies have (approximately)
the same study sizes~\cite{Cochran1954,Bliss1952}, but in medical
sciences it is frequently ignored~\cite{Borenstein2009}.}~\cite{Cochran1954}. The aggregated data is typically constructed from
individual data (although a researcher may not have access to this
individual data). The aggregated data is then used to determine a
pooled estimate of the main effect size and to calculate other aspects
related to the quality of the meta-analysis (e.g., measure of heterogeneity,
publication bias, sensitivity analyses).

To investigate the performance of meta-analysis methods, researchers
often resort to simulation of the aggregated data $(D_{i},S_{i},df_{i})$
directly, making certain distributional assumptions~\cite{Brockwell2001, Chung2013, Rukhin2013, Ning2017}. The distribution of $D_i$ is typically assumed normal and the standard error $S_i$ is often related to a chi-square distribution.~\cite{jackson_when_2018} While these distributional assumptions may be plausible,
the choices are not properly supported or defended by an underlying
statistical model on the individual data that would generate the aggregated
data, making it difficult to understand if the simulation model for the aggregated statistics represents the practice appropriately. On the other hand, when individual data is being simulated~\cite{Stanley2008, Stanley2014, Alinaghi2018} there
is limited discussion on the distributional characteristics of the aggregated data $(D_i, S_i, df_i)$. This makes it difficult to compare simulations on an individual level with simulations conducted at the aggregated level and it is hard to verify the compliance of the distribution of the aggregated statistics in practice with the induced distributional assumptions of the individual data. Furthermore, research papers often choose but one simulation approach and it is typically unknown how well the conclusions of this single simulation model would hold for other choices of simulation. 

This paper discusses a (fixed and random effects) heteroscedastic
mixed effects model for a continuous outcome at an individual level
where two groups (e.g., treatments) are being compared. The model
assumes three forms of heterogeneity. The first form is a bivariate
random effect on the two mean outcomes across studies, indicating
heterogeneity in outcome level between individuals across studies.
The second form is a fixed heteroscedastic residual error for the
two groups within a study, implying that inter-individual variability
within studies can depend on other variables (like treatment). The
third form is a random heteroscedastic residual error across studies,
which would represent that inter-individual variability can also change
with studies due to the selection of more homogeneous or heterogeneous
participants. We believe that this model has not been discussed for
meta-analysis purposes so far. Based on this model we will discuss
distributional properties of the aggregated data $(D_{i},S_{i},df_{i})$,
which is useful to be able to simulate study data at an aggregated
level directly.

These distributional properties will be compared with similar distributional
properties of certain other aggregated data simulation models from
literature to investigate the plausibility of these choices. Secondly,
we will conduct a simulation study to investigate the impact of different
distributional choices on some well-known meta-analysis approaches.
We selected the very popular pooled estimation method of DerSimonian
and Laird for the random effects meta-analysis model \cite{DerSimonian1986}, the popular Trim \& Fill method~\cite{Duval2000, Duval2000a} for adjustments of publication bias
in meta-analysis practice~\cite{Dadi2020, Liu2020, Torous2020}, and the PET-PEESE approach~\cite{Stanley2008, Stanley2014} also for adjustment of publication
bias. The PET-PEESE method has been popular in at least the economics
field~\cite{Alinaghi2018}.

In section 2 we describe a selection of aggregated data simulation models we have encountered in literature.
In section 3, we propose a statistical model for individual participant
data and determine distributional properties of the corresponding aggregated
statistics. We also provide a simulation procedure to simulate aggregated
data according to this model. We discuss the differences
between our approach with the selected models from the literature as well. The fourth section presents a simulation study where
the aggregated data simulation models are being compared. We also
provide the simulation settings and a selection model for publication
bias used in literature as well. Then finally we provide a discussion
of the results in Section 6.

\section{Simulation of Aggregated Data in Literature\label{sec:Simulation-Literature}}

In this section we will provide an overview of a few simulation models
that we have encountered in literature to support investigations of
the performance of meta-analysis approaches. The simulation models used in literature all uses the well-known random
effects model~\cite{DerSimonian1986, Hardy1996}
for the study effect size $D_{i}$ at the aggregated level that is
given by
\begin{equation}
D_{i}=\theta+U_{i}+\varepsilon_{i},\label{eq:ADMA-random-effect}
\end{equation}
with $\theta$ the mean study effect of interest, $U_{i}\sim\mathcal{N}(0,\tau^{2})$
a random effect for study $i$ to accommodate study heterogeneity
on the treatment effect (i.e., $\theta+U_{i}$ is the effect size
for study $i$), $\varepsilon_{i}\sim\mathcal{N}(0,\sigma_{i}^{2})$
the residual for study $i$, and $U_{i}$ and $\varepsilon_{i}$ independent. Thus, $D_i$ is created by drawing $U_i$ and $\varepsilon_i$ for study $i$.

Simulations at the aggregated level also have to generate estimators $S_i^2$ for the variance component $\sigma_i^2$ of the residual. Differences between the aggregated data simulation models in literature
are typically determined by the choice of distribution function for
the standard deviation $S_{i}$. The choice of distributions have
been the central chi-square for $S_{i}^{2}$ in Sidik and Jonkman~\cite{Sidik2002} and Brockwell and Gordon~\cite{Brockwell2001}, the non-central chi-square
for $S_{i}^{2}$ in Ning \textit{et al.}~\cite{Ning2017}, and the gamma distribution
for $S_{i}$ in Duval and Tweedie~\cite{Duval2000a}. All these aggregated data simulation models assume that $\sigma_i^2=\mathsf{VAR}(\varepsilon_i|S_i^2)=S_i^2$, indicating that $S_i^2$ is drawn first before the study effect $D_i$ is simulated. Furthermore, the size of a study, which may be represented by the degrees of freedom $df_i$ of $S_i^2$, is not considered as a separate measure by these simulation models.

\subsection{Central Chi-Square Distributed $S_{i}^{2}$}

For the simulation of log odds ratios in an aggregated data meta-analysis, Brockwell
and Gordon~\cite{Brockwell2001} and Sidik and Jonkman~\cite{Sidik2002} simulated the log odds
ratio $D_{i}$ according to model (\ref{eq:ADMA-random-effect}). Here $S_i^2$ is selected to be equal to $S_{i}^{2}=0.25\chi_{i}^{2}$,
with $\chi_{i}^{2}$ a chi-square distributed random variable having
only one degrees of freedom (thus effectively choosing $df_{i}=1$). They also restricted the
value of $S_{i}^{2}$ to be within $(0.009,0.6)$ to conform with
a typical distribution of the variances of log odds-ratios in practice~\cite{Brockwell2001, Sidik2002}. For mean differences
we do not think that we should implement these variance restrictions.
In their comparisons of several meta-analysis methods of pooling log
odd ratios they did not make use of the $df_{i}$ in any way. They
selected a mean study effect of $\theta=0.5$ and evaluated eleven
different values for the heterogeneity $\tau^{2}$ in the range of
$0$ to $0.1$.

\subsection{Non-Central Chi-Square Distributed $S_{i}^{2}$}

For an investigation of a publication bias method, Ning et al.,~\cite{Ning2017}
simulated the study effect size $D_{i}$ according to model (\ref{eq:ADMA-random-effect})
with condition $\mathsf{VAR}(\varepsilon_{i}|S_{i}^{2})=S_{i}^{2}$.
And to create the variance
$S_{i}^{2}$, they draw a normally distributed random variable $Z_{i}\sim N(0.25,0.5)$
and calculated the variance by $S_{i}^{2}=Z_{i}^{2}$. The distribution
of this variance is equal to a non-central chi-square distribution
with non-centrality parameter equal to $1/8$ ($=(0.25)^{2}/0.5$).
They did not implement the degrees of freedom $df_{i}$.The mean effect size was taken equal to $\theta=0.4$ and the heterogeneity
was selected equal to $\tau=0.5$ and $\tau=1$.

\subsection{Gamma Distributed $S_{i}$}

For an investigation of the Trim \& Fill method for publication bias,
Duval and Tweedie~\cite{Duval2000, Duval2000a} simulated the study effect size $D_{i}$
according to model (\ref{eq:ADMA-random-effect}) with a standard error $S_{i}$ drawn from a Gamma
distribution with parameters $3$ and $1/9$ (i.e., $S_{i}\sim\Gamma(3,1/9)$).
Their argument for this choice is that it provides a more typical
funnel plot than a uniform distribution for $S_{i}$, which was applied
by Light and Pillemer~\cite{Light1986}. They did not specify which parametrization
of the gamma distribution they selected, but we guess that the mean
of $S_{i}$ would be equal to $1/3$ ($=3\times1/9$) and with a variance
equal to $1/27$ ($=3\times(1/9)^{2}$), considering the values for
precision $S_{i}^{-1}$ they used in their funnel plots. They assumed
that both the mean study effect $\theta$ and heterogeneity $\tau^{2}$
were equal to zero.

\section{Heteroscedastic Mixed Effects Model\label{sec:Statistical-Model}}

Different data-generating mechanisms in simulation studies for meta-analysis may potentially affect the results when comparing the performances of different meta-analysis methods. In this section, we will describe an individual participant data (IPD) simulation model which allows treatment heterogeneity and heteroscedasticity across treatment arms and across trials.  Furthermore, we will show that the
distributional properties of the aggregated data $D_{i}$, $S_{i}^{2}$,
and $df_{i}$ for a meta-analysis that we obtained from our heterogeneous
and heteroscedastic IPD model deviates from what is normally used in literature.

Let $Y_{ijk}$ be a continuous outcome variable for individual $k=1,2,...,n_{ij}$
in treatment group $j=0,1$ for study $i=1,2,...,m$. Treatment $j=0$
represents the control group and treatment $j=1$ is the treatment
under investigation. Note that we use treatment here, but it could
be in principle any binary variable that splits the data in two groups
(e.g., sex, symptom, disease). The statistical model is described
as follows:
\begin{equation}
Y_{ijk}=\mu+\beta_{j}+U_{ij}+\varepsilon_{ijk}\label{eq:IPD model}
\end{equation}
with $\mu$ the mean response at the control group, $\beta_{0}=0$
for identifiability purposes, $\beta_{1}$ the mean treatment effect
and typically the (association) parameter of interest, $U_{ij}$ the
(latent) study heterogeneity for treatment group $j$, and $\varepsilon_{ijk}$
a potentially heteroscedastic residual. We will assume that the random
elements are normally distributed, such that
\begin{equation}
\begin{array}{rcl}
\varepsilon_{ijk}|V_{i} & \sim & \mathcal{N}(0,\sigma_{j}^{2}\exp\left\{ V_{i}\right\} ),\\
\left(U_{i0},U_{i1},V_{i}\right)^{T} & \sim & \mathcal{N}(\boldsymbol{0},\boldsymbol{\Sigma}),
\end{array}\label{eq:IPD assumptions}
\end{equation}
with $V_{i}$ a (latent) random heteroscedastic study effect on the
residuals in (\ref{eq:IPD model}), with $\boldsymbol{\Sigma}$ given
by
\[
\boldsymbol{\Sigma}=\left(\begin{array}{ccc}
\tau_{0}^{2} & \rho_{01}\tau_{0}\tau_{1} & \rho_{02}\tau_{0}\tau_{2}\\
\rho_{01}\tau_{0}\tau_{1} & \tau_{1}^{2} & \rho_{12}\tau_{1}\tau_{2}\\
\rho_{02}\tau_{0}\tau_{2} & \rho_{12}\tau_{1}\tau_{2} & \tau_{2}^{2}
\end{array}\right),
\]
and $\varepsilon_{ijk}$ and $(U_{i0},U_{i1})$ being independent
conditionally on $V_{i}$. It should be noted that our assumptions
of normality are not uncommon for the analysis of clinical trials
and observational studies~\cite{Molenberghs2007, Rosenbaum2002}. Furthermore, heteroscedastic hierarchical models defined by
(\ref{eq:IPD model}) and (\ref{eq:IPD assumptions}) have been discussed
in literature as well~\cite{Davidian1987, Quintero2017}, but not in the context of meta-analysis.

Our model describes study heterogeneity ($U_{ij}$) on the mean response,
indicating that the outcome level may vary with studies, but that
this form of heterogeneity is not the same for both treatment groups.
This may not be unrealistic when the treatment group would fully cure
a disease but the control group does not, assuming that variability
within a population is larger when it is composed of healthy and diseased
individuals. It is this form of heterogeneity that leads to the (well-known)
random effects meta-analysis model~\eqref{eq:ADMA-random-effect} at the aggregated level. When the heterogeneity would be identical
for both treatment groups ($\rho_{01}=1$ and $\tau_{1}=\tau_{2}$),
we may only observe heterogeneity at the individual level data for
each treatment group, but not necessarily at the aggregated study
effect (see Section \ref{subsec:Aggregated-Data}).

Furthermore, we also assume that the residual variability, which represents
inter-individual variability, is treatment dependent (i.e., $\sigma_{0}$
and $\sigma_{1}$ may be different). This is not uncommon in several
areas of medicine (e.g., hypertension treatment) where treatment is
not just affecting the level of the outcome in a population, but also
its variability. Our model also describes a residual heteroscedasticity
that is study dependent ($V_{i}$). This could be related to the selection
of participants in a study, leading to a study with either more homogeneous
or less homogeneous participants. For instance, a clinical trial with
mostly men at a specific age group would show less inter-individual
variability than a trial with both sexes and a larger variety in age
groups. Combining these different trials in a meta-analysis results
in a random residual heteroscedasticity. Here we assume that we do
not have any complementary data to investigate or eliminate such heteroscedasticity,
implying the necessity of $V_{i}$ in model (\ref{eq:IPD model}).

Model (\ref{eq:IPD model}) can be rewritten into $Y_{ijk}=\mu+\beta_{j}+U_{ij}+\sigma_{j}Z_{ijk}\exp\left\{ 0.5V_{i}\right\}$,
with $Z_{ijk}$ i.i.d.\ standard normally distributed residuals (i.e.,
$Z_{ijk}\sim\mathcal{N}(0,1)$), which are independent of the random
effects $U_{i0}$, $U_{i1}$, and $V_{i}$. This formulation demonstrates
immediately that the selected model (\ref{eq:IPD model}) with assumptions
(\ref{eq:IPD assumptions}) has introduced a correlation between the
residuals $\sigma_{j}Z_{ijk}\exp\left\{ 0.5V_{i}\right\}$ and the random location effects $(U_{i0}, U_{i1})$. In case we choose $\rho_{02}=\rho_{12}=0$,
the error structure in (\ref{eq:IPD model}) and the study treatment
heterogeneity $U_{ij}$ will be independent. It also shows that the
marginal distribution of $Y_{ijk}$ is not normal anymore, due to
the product of a normal and log normal random variable in the residual, unless $\tau_{2}=0$
of course.

Finally, the distribution of $U_{i1}-U_{i0}$ and $V_{i}$ is bivariate
normally distributed with mean $\boldsymbol{0}$ and a variance-covariance
matrix given by
\[
\left(\begin{array}{cc}
\tau_{0}^{2}-2\rho_{01}\tau_{0}\tau_{1}+\tau_{1}^{2} & \tau_{2}[\rho_{12}\tau_{1}-\rho_{02}\tau_{0}]\\
\tau_{2}[\rho_{12}\tau_{1}-\rho_{02}\tau_{0}] & \tau_{2}^{2}
\end{array}\right).
\]
The difference $U_{i1}-U_{i0}$ is important in Section \ref{subsec:Aggregated-Data}
when we determine and discuss the study effect size $D_i$. The correlation
coefficient $\rho$ between $U_{i1}-U_{i0}$ and $V_{i}$ is 
equal to $\rho=[\rho_{12}\tau_{1}-\rho_{02}\tau_{0}]/\sqrt{\tau_{0}^{2}-2\rho_{01}\tau_{0}\tau_{1}+\tau_{1}^{2}}$.
We will demonstrate that the study effect size $D_i$ and its corresponding
standard error $S_i$ will always be dependent even in the absence of correlation ($\rho=0$) between $U_{i1}-U_{i0}$ and $V_i$ (i.e., $\rho_{02}\tau_{0}=\rho_{12}\tau_{1}$),
unless $\tau_{2}=0$ of course.

\subsection{Aggregated Effects Sizes and Standard Errors\label{subsec:Aggregated-Data}}

A popular effect size used in meta-analysis calculated for each study
is the (raw) mean difference between the two treatment groups\footnote{The raw mean difference is more common to use when all studies measure
the outcome on the same scale (e.g., hypertension treatment for blood
pressure). In the educational or social sciences, it is common to
have scales that may vary with study and it would be more appropriate
to divide the mean difference by a standard deviation to create a
standardized measure of effect size in the form of Cohen's $d$~\cite{McGaw1980}. We will assume a common scale.}, i.e.,
\begin{equation}
D_{i}=\bar{Y}_{i1}-\bar{Y}_{i0},
\end{equation}
with $\bar{Y}_{ij}=\sum_{k=1}^{n_{ij}}Y_{ijk}/n_{ij}$ the average
response for study $i$ at treatment $j$. Using model (\ref{eq:IPD model}),
we can rewrite the effect size into
\begin{equation}
D_{i}=\beta_{1}+U_{i1}-U_{i0}+\exp\left\{ 0.5V_{i}\right\} \left[\sigma_{1}\bar{Z}_{i1}-\sigma_{0}\bar{Z}_{i0}\right],\label{eq:Effect_size}
\end{equation}
with $\bar{Z}_{ij}=\sum_{k=1}^{n_{ij}}Z_{ijk}/n_{ij}$ the average
standardized residuals for treatment $j$ in study $i$.  Clearly, 
model (\ref{eq:Effect_size}) is identical to model (\ref{eq:ADMA-random-effect}), with $\theta=\beta_{1}$, $U_{i}=U_{i1}-U_{i0}$
and $\varepsilon_{i}=\exp\left\{ 0.5V_{i}\right\} \left[\tau_{1}\bar{Z}_{i1}-\tau_{0}\bar{Z}_{i0}\right]$,
but the distributional assumptions on normality and independence only
agree with our setting when the random heteroscedasticity vanishes
(i.e., $\tau_{2}=0$). Under our model~\eqref{eq:Effect_size}, the variances $\tau^{2}$ and $\sigma_{i}^{2}$
in model (\ref{eq:ADMA-random-effect}) would become equal to $\tau^{2}=\tau_{0}^{2}-2\rho_{01}\tau_{0}\tau_{1}+\tau_{1}^{2}$
and $\sigma_{i}^{2}=\exp\{\tau_{2}^{2}/2\}[\sigma_{0}^{2}/n_{i0}+\sigma_{1}^{2}/n_{i1}]$,
respectively. The difference $U_{i1}-U_{i0}\sim N(0,\tau_{0}^{2}-2\rho_{01}\tau_{0}\tau_{1}+\tau_{1}^{2})$
represents the study heterogeneity of effect sizes, but it disappears
when $\rho_{01}=1$ and $\tau_{0}=\tau_{1}$.

The conditional distribution of the effect size $D_{i}$, given the
three random effects $U_{i0}$, $U_{i1}$, and $V_{i}$, is given
by a normal distribution with mean $\beta_{1}+U_{i1}-U_{i0}$ and
variance $\exp\left\{ V_{i}\right\} \left[\sigma_{0}^{2}/n_{i0}+\sigma_{1}^{2}/n_{i1}\right]$.
A larger study will demonstrate a smaller residual variance, since
the sample sizes $n_{i0}$ and $n_{i1}$, which will then be smaller
in other studies, will reduce the residual variance, but it could
potentially be compensated by a larger inter-individual variability
($V_{i}>0$). However, we do not assume a direct relation between
a positive value for $V_{i}$ and the sample size $n_{i0}$ and $n_{i1}$
(i.e., $P(V_{i}>0|n_{i0},n_{i1})=P(V_{i}>0)$). The conditional distribution
of $D_{i}$ given only $V_{i}$ is also given by a normal distribution,
but now with mean and variance given by
\begin{equation}
\begin{array}{rcl}
\mathbb{E}(D_{i}|V_{i}) & = & \beta_{1}+\tau_{2}^{-1}[\rho_{12}\tau_{1}-\rho_{02}\tau_{0}]V_{i},\\
\mathbb{\mathsf{VAR}}(D_{i}|V_{i}) & = & (1-\rho_{02}^{2})\tau_{0}^{2}+2(\rho_{02}\rho_{12}-\rho_{01})\tau_{0}\tau_{1}+(1-\rho_{12}^{2})\tau_{1}^{2}\\
 &  & \qquad+\exp\left\{ V_{i}\right\} \left[\sigma_{0}^{2}/n_{i0}+\sigma_{1}^{2}/n_{i1}\right].
\end{array}\label{eq:Conditional_Di}
\end{equation}
Finally, the marginal distribution function for $D_{i}$ is less tractable
when $V_{i}$ is non-degenerative, since the distribution of the product
of a log normal and a normal distributed random variable (i.e., $\exp\left\{ 0.5V_{i}\right\} \left[\sigma_{0}\bar{Z}_{i1}-\sigma_{1}\bar{Z}_{i0}\right]$)
is unknown whether they are dependently or independently distributed
(as we already mentioned for $Y_{ijk}$). However, we can determine
the mean and variance of $D_{i}$ and they are given by
\begin{equation}
\begin{array}{rcl}
\mathbb{E}(D_{i}) & = & \beta_{1},\\
\mathbb{\mathsf{VAR}}(D_{i}) & = & \tau_{0}^{2}-2\rho_{01}\tau_{0}\tau_{1}+\tau_{1}^{2}+\exp\left\{ \tau_{2}^{2}/2\right\} [\sigma_{0}^{2}/n_{i0}+\sigma_{1}^{2}/n_{i1}].
\end{array}\label{eq:Moments_Di}
\end{equation}
Note that correlation coefficient $\rho=[\rho_{12}\tau_{1}-\rho_{02}\tau_{0}]/\sqrt{\tau_{0}^{2}-2\rho_{01}\tau_{0}\tau_{1}+\tau_{1}^{2}}$
does not play a role in the variance of $D_{i}$, due to the independence
of $U_{i1}-U_{i0}$ and $\sigma_{1}\bar{Z}_{i1}-\sigma_{0}\bar{Z}_{i0}$.
Thus the variance is just the sum of the variance of the heterogeneity
$U_{i1}-U_{i0}$ and the variance of the residual $\exp\left\{ 0.5V_{i}\right\} \left[\sigma_{1}\bar{Z}_{i1}-\sigma_{0}\bar{Z}_{i0}\right]$.

The estimated standard error $S_{i}$ for the effect size $D_{i}$
within one study (thus being unaware of any possible random effect
$U_{i0}-U_{i1}$ across studies) can be estimated in different ways~\cite{Borenstein2009}. If we do not want to make any
assumptions on the equality of residual variances for the two treatment
groups in model (\ref{eq:IPD model}), it is most reasonable to estimate
the standard error $S_{i}$ by $S_{i}^{2}=S_{i0}^{2}/n_{i0}+S_{i1}^{2}/n_{i1}$,
with $S_{ij}^{2}=\sum_{k=1}^{n_{ij}}(Y_{ijk}-\bar{Y}_{ij})^{2}/(n_{ij}-1)$
the sample variance for treatment $j$ in study $i$\footnote{Under assumption of homogeneous residual variances within studies
($\sigma_{0}^{2}=\sigma_{1}^{2})$, we may choose the pooled variance
$S_{i}^{2}=[n_{i0}^{-1}+n_{i1}^{-1}]\left[(n_{i0}-1)S_{i0}^{2}+(n_{i1}-1)S_{i1}^{2}\right]/\left[n_{i0}+n_{i1}-2\right]$,
instead of $S_{i}^{2}=S_{i0}^{2}/n_{i0}+S_{i1}^{2}/n_{i1}$. Derivations
on this pooled variance using our model (\ref{eq:IPD model}) with
assumptions (\ref{eq:IPD assumptions}) can be implemented in the
same way as the derivations we will apply to $S_{i}^{2}=S_{i0}^{2}/n_{i0}+S_{i1}^{2}/n_{i1}$.
These derivations are actually somewhat simpler than for our choice
of $S_{i}^{2}=S_{i0}^{2}/n_{i0}+S_{i1}^{2}/n_{i1}$ and it leads to
a central chi-square distribution with $n_{i0}+n_{i1}-2$ degrees
of freedom conditionally on $V_{i}$ and when properly scaled.}. Using model (\ref{eq:IPD model}), we can rewrite this sample variance
$S_{ij}^{2}$ into $\sigma_{j}^{2}\exp\left\{ V_{i}\right\} \chi_{ij}^{2}/(n_{ij}-1)$,
with $\chi_{ij}^{2}=\sum_{k=1}^{n_{ij}}(Z_{ijk}-\bar{Z}_{ij})^{2}$
a chi-square distributed random variable with $n_{ij}-1$ degrees
of freedom. Thus the variance $S_{i}^{2}$ is now distributed as
\begin{equation}
S_{i}^{2}=\exp\left\{ V_{i}\right\} [\sigma_{0}^{2}\chi_{i0}^{2}/(n_{i0}(n_{i0}-1))+\sigma_{1}^{2}\chi_{i1}^{2}/(n_{i1}(n_{i1}-1))],\label{eq:Standard_Error}
\end{equation}
with an expectation equal to $\mathbb{E}(S_{i}^{2})=\exp\left\{ \tau_{2}^{2}/2\right\} [\sigma_{0}^{2}/n_{i0}+\sigma_{1}^{2}/n_{i1}]$
and a variance equal to
\begin{equation}
\begin{array}{l}
\mathsf{VAR}(S_{i}^{2})=2\exp\left\{ 2\tau_{2}^{2}\right\} \left[\sigma_{0}^{4}/(n_{i0}^{2}(n_{i0}-1))+\sigma_{1}^{4}/(n_{i1}^{2}(n_{i1}-1))\right]\\
\qquad\qquad\qquad\qquad\qquad+\exp\left\{ \tau_{2}^{2}\right\} \left(\exp\left\{ \tau_{2}^{2}\right\} -1\right)\left[\sigma_{0}^{2}/n_{i0}+\sigma_{1}^{2}/n_{i1}\right]^{2}.
\end{array}\label{eq:Variance_Si}
\end{equation}

The random variable $\sigma_{0}^{2}\chi_{i0}^{2}/(n_{i0}(n_{i0}-1))+\sigma_{1}^{2}\chi_{i1}^{2}/(n_{i1}(n_{i1}-1))$
is approximately distributed as $[\sigma_{0}^{2}/n_{i0}+\sigma_{1}^{2}/n_{i1}]\chi_{df_{i}}^{2}/df_{i}$,
with $\chi_{df_{i}}^{2}$ a chi-square distributed random variable
with $df_{i}$ degrees of freedom. The number of degrees of freedom
is given by~\cite{Satterthwaite1946, Heuvel2010}
\begin{equation}
df_{i}=\dfrac{\left[\sigma_{0}^{2}\chi_{i0}^{2}/(n_{i0}(n_{i0}-1))+\sigma_{1}^{2}\chi_{i1}^{2}/(n_{i1}(n_{i1}-1))\right]^{2}}{\sigma_{0}^{4}\chi_{i0}^{4}/\left[n_{i0}^{2}(n_{i0}-1)^{3}\right]+\sigma_{1}^{4}\chi_{i1}^{4}/\left[n_{i1}^{2}(n_{i1}-1)^{3}\right]}.\label{eq:Satterthwaite}
\end{equation}
This number of degrees of freedom is bounded from above by $n_{i0}+n_{i1}-2$
and from below by $\min\left\{ n_{i0}-1,n_{i1}-1\right\} $~\cite{Heuvel2010}. In case the sample sizes are equal ($n_{i0}=n_{i1}$)
and the residual variances are equal ($\sigma_{0}^{2}=\sigma_{1}^{2}$),
the random variable $\chi_{df_{i}}^{2}$ becomes chi-square with $df_{i}=n_{i0}+n_{i1}-2$
degrees of freedom. The distribution of $S_{i}^{2}$ in the general
setting (\ref{eq:Standard_Error}) using model (\ref{eq:IPD model})
with assumptions (\ref{eq:IPD assumptions}) is clearly intractable,
but becomes approximately equal to a chi-square distribution when
$V_{i}$ becomes degenerate in zero.

The distribution of $S_i^2$ in (\ref{eq:Standard_Error}) seems clearly different from those used in literature, since it is the distribution of a random variable that is a product of a lognormally distributed random variable and a weighted average of two chi-square distributed variables, where the weights depend on the sample sizes. Even if there is no residual heteroscedasticity ($\sigma_0=\sigma_1$ and $\tau_2=0$) and the sample sizes $n_{i0}$ and $n_{i1}$ are equal, $S_i^2$ has a central chi-square distribution with $df_i=n_{i0}+n_{i1}-2$ degrees of freedom and still deviates from the distributions for $S_i^2$ in Section~\ref{sec:Simulation-Literature}. However, we have only studied the distribution function of $S_i^2$ conditionally on the sample sizes $n_{i0}$ and $n_{i1}$. In meta-analysis these sample sizes are typically known and could therefore be used in the analysis of a meta-analysis (e.g., by calculating appropriate degrees of freedom $df_i$), but for the central chi-square, the non-central chi-square, and gamma distributions in Section~\ref{sec:Simulation-Literature}, these sample sizes are not being mentioned and, in a way, they can be viewed as distribution functions where the sample sizes have been integrated out (i.e., they essentially describe marginal distribution functions of $S_i^2$). Indeed, we do not know if there exist distribution functions for $n_{i0}$ and $n_{i1}$ that would make the marginal distribution of $S_i^2$ in (\ref{eq:Effect_size}) equal to any of the distribution functions in Section~\ref{sec:Simulation-Literature}. 

Note that $D_{i}$ and $S_{i}^{2}$ are independent conditionally
on $V_{i}$, due to independence of the random variables $\chi_{i0}^{2}$,
$\chi_{i1}^{2}$, $\bar{Z}_{i1}$, $\bar{Z}_{i1}$, and $U_{i0}-U_{i1}$.
Here we have used the assumption of normality to obtain independence
between the mean $\bar{Z}_{ij}$ and variance $S_{ij}^{2}$ (conditionally
on $V_{i}$). Using this independence, (\ref{eq:Effect_size}), and
(\ref{eq:Standard_Error}), the joint distribution of $D_{i}$ and
$S_{i}^{2}$ is now a mixture of the product of the two conditional
distribution functions for $D_{i}|V_{i}$ and $S_{i}^{2}|V_{i}$.
This implies that $D_{i}$ and $S_{i}^{2}$ are never independent,
irrespective of the value of $\rho$, unless $V_{i}$ is degenerated.
The covariance between $D_{i}$ and $S_{i}^{2}$ is equal to
\begin{equation}
\begin{array}{rcl}
\mathtt{\mathsf{COV}}(D_{i},S_{i}^{2}) & = & \mathbb{E}\left[(D_{i}-\beta_{1})S_{i}^{2}\right],\\
 & = & \tau_{2}^{-1}[\rho_{12}\tau_{1}-\rho_{02}\tau_{0}]\mathbb{E}\left[V_{i}\exp\left\{ V_{i}\right\} \right][\sigma_{0}^{2}/n_{i0}+\sigma_{1}^{2}/n_{i1}],\\
 & = & \tau_{2}[\rho_{12}\tau_{1}-\rho_{02}\tau_{0}]\exp\left\{ \tau_{2}^{2}/2\right\} [\sigma_{0}^{2}/n_{i0}+\sigma_{1}^{2}/n_{i1}],
\end{array}\label{eq:Cov_Di_Si}
\end{equation}
and depends on the covariance of $U_{i0}-U_{i1}$ and $V_{i}$ (see
Section \ref{sec:Statistical-Model}). The covariance in (\ref{eq:Cov_Di_Si})
vanishes when $\rho_{12}\tau_{1}-\rho_{02}\tau_{0}=0$ even though they remain dependent. This occurs when the heterogeneity and heteroscedasticity are unrelated.
Using the variance of $D_{i}$ in (\ref{eq:Moments_Di}) and the variance
of $S_{i}^{2}$ in (\ref{eq:Variance_Si}) the correlation between
$D_{i}$ and $S_{i}^{2}$ can be obtained as well. Without heteroscedasticity, we would also obtain a
variance $S_{i}^{2}$ that is related to a central chi-square distribution
when we would change our estimator $S_{i}^{2}$ into $S_{i}^{2}=S_{p}^{2}[n_{i0}^{-1}+n_{i1}^{-1}]$,
with $S_{p}^{2}=\left[(n_{i0}-1)S_{i0}^{2}+(n_{i1}-1)S_{i1}^{2}\right]/\left[n_{i0}+n_{i1}-2\right]$
the pooled variance within study $i$.

\subsection{Procedure for Simulating Aggregated Data\label{subsec:Simulation-Procedure}}

One approach is to simulate individual data via model (\ref{eq:IPD model})
for all $m$ studies and then calculate all the necessary aggregated
data statistics $D_{i}$, $S_{i}^{2}$, and $df_{i}$ per study. This
requires input values for the sample sizes $m,$ $n_{i0}$, and $n_{i1}$
and all parameters in model (\ref{eq:IPD model}) and (\ref{eq:IPD assumptions}).
Alternatively, we could simulate the aggregated data statistics directly.
Then we only need to set values for the following parameters: sample
sizes $m$, $n_{i0}$, and $n_{i1}$, treatment effect $\beta_{1}$,
variance component $\tau^{2}=\tau_{0}^{2}-2\rho_{01}\tau_{0}\tau_{1}+\tau_{1}^{2}$
for a heterogeneous treatment effect, variance component $\tau_{2}^{2}$
for a random heteroscedasticity, correlation coefficient $\rho=\tau^{-1}[\rho_{12}\tau_{1}-\rho_{02}\tau_{0}]$
between heterogeneity and heteroscedasticity, and residual variances
$\sigma_{0}^{2}$ and $\sigma_{1}^{2}$. Here we describe the aggregated
approach.

Since studies often have different sample sizes we first need to choose
a distribution for the sample sizes $n_{i0}$ and $n_{i1}$. Although
alternative approaches could be used, we first draw one sample size
$n_{i}$ for study $i$ using a mixture of a Poisson-Gamma distribution
(i.e., an overdispersed Poisson). We draw a random variable $\gamma_{i}$
from a gamma distribution $\Gamma\left(a,b\right)$ with shape and
scale parameters $a$ and $b$, respectively, (having mean $a/b$
and variance $a/b^{2}$) and then conditionally on this result we
draw a sample size $n_{i}$ from a Poisson distribution with mean
parameter $\lambda_{i}=\lambda_{0}\exp\left\{ \gamma_{i}\right\} $.
Then we split the sample size $n_{i}$ into $n_{i0}$ and $n_{i1}$
using a Binomial distribution. We assumed that the distribution of
$n_{i0}|n_{i}\sim\mathrm{Bin}(n_{i},p)$, with $p$ a proportion which
can take different values to create potential imbalances in sample
sizes for the two groups.

Since $D_{i}$ and $S_{i}^{2}$ are independently distributed given
$V_{i}$ and $\chi_{i0}^{2}$, $\chi_{i1}^{2}$, and $V_{i}$ are
also mutually independent, we could now independently draw $\chi_{i0}^{2}$,
$\chi_{i1}^{2}$, and $V_{i}$ for study $i$ using the chi-square
distributions with $n_{i0}-1$ and $n_{i1}-1$ degrees of freedom
and the normal distribution $\mathcal{N}(0,\tau_{2}^{2})$, respectively.
Then $S_{i}^{2}$ can be calculated directly using equation (\ref{eq:Standard_Error})
and $D_{i}$ can be directly drawn from a normal distribution using
the mean and variance in (\ref{eq:Conditional_Di}). Note that the
mean can be rewritten into $\beta_{1}+\tau_{2}^{-1}\tau\rho V_{i}$
and the variance into $\tau^{2}(1-\rho^{2})+\exp\{V_{i}\}\sigma_{i}^{2}$,
with $\sigma_{i}^{2}=\sigma_{0}^{2}/n_{i0}+\sigma_{1}^{2}/n_{i1}$,
using only parameters at the aggregated level. The degrees of freedom
can be calculated from (\ref{eq:Satterthwaite}) if this would be
needed.

It should be noted that we need two chi-square distributed random
variables $\chi_{i0}^{2}$ and $\chi_{i1}^{2}$ to be able to obtain
the exact distribution of $S_{i}^{2}$ given $V_{i}$ and to calculate
the degrees of freedom $df_{i}$ properly. Although we could argue
that one chi-square distributed random variable $\chi_{df_{i}}$ would
be appropriate based on Satterthwaite approximation theory, it is
not that easy to simulate since we need to know the degrees of freedom
$df_{i}$ in (\ref{eq:Satterthwaite}). Indeed, this degrees of freedom
is based on the data and we do not know its distribution, making a
simulation of one chi-square distributed random variable more complicated.

There is another important difference between our aggregated data
simulation and the simulations performed in literature. Our simulation
would draw $D_{i}$ independently from $S_{i}^{2}$ when conditioned
on the random heteroscedastic effect $V_{i}$. Thus dependency between
$D_{i}$ and $S_{i}^{2}$ is indirectly created through the random
variable $V_{i}$ and with a covariance given in (\ref{eq:Cov_Di_Si}).
In literature, the dependency between $D_{i}$ and $S_{i}^{2}$ is
derived directly through $S_{i}^{2}$, since the residuals in (\ref{eq:ADMA-random-effect})
are typically being drawn from a normal distribution with mean zero
and variance $\sigma_{i}^{2}=S_{i}^{2}$. Although $D_{i}$ and $S_{i}^{2}$
are dependent in this way, the covariance between $D_{i}$ and $S_{i}^{2}$
is still equal to zero and therefore differs from our setting. We
would also obtain a covariance of zero when either $V_{i}$ is degenerated
(but then $D_{i}$ and $S_{i}^{2}$ are independent) or otherwise
when $U_{i1}-U_{i0}$ and $V_{i}$ are independent. We believe that
the assumption $\mathsf{VAR}(\varepsilon_{i}|S_{i}^{2})=S_{i}^{2}$
in model (\ref{eq:ADMA-random-effect}) is highly unlikely in reality,
as our IPD model shows, although we are aware that both estimation
approaches of DerSimonian and Laird~\cite{DerSimonian1986}
and Hardy and Thompson~\cite{Hardy1996} assumed that the
standard error $S_{i}$ was the true standard deviation of the residual
$\varepsilon_{i}$ in model (\ref{eq:ADMA-random-effect}). However,
this does not imply that we should also simulate according to this
assumption.

\section{Simulation Study}

\begin{figure}
\begin{subfigure}{.5\textwidth}
  \centering
  \includegraphics[width=.8\linewidth]{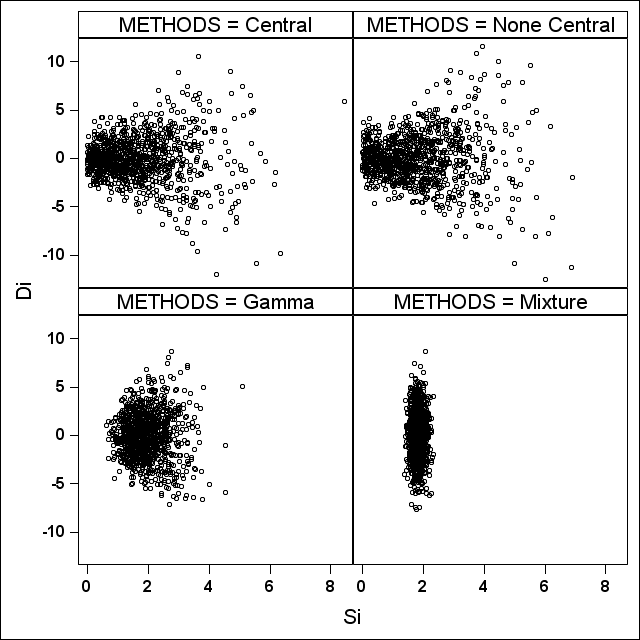}  
  \caption{$\tau_2^2=0$ and $\rho=0$}
  \label{fig:sub-first}
\end{subfigure}
\begin{subfigure}{.5\textwidth}
  \centering
  \includegraphics[width=.8\linewidth]{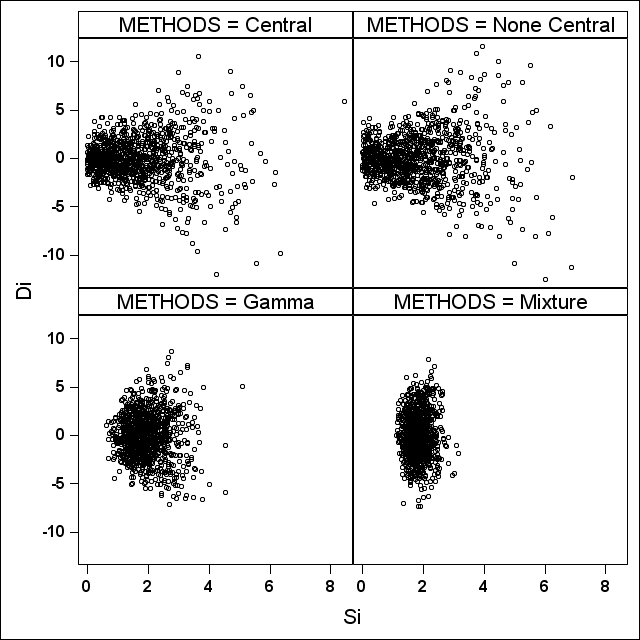}  
  \caption{$\tau_2^2=\log(2)$ and $\rho=0$}
  \label{fig:sub-second}
\end{subfigure}
\newline
\begin{subfigure}{.5\textwidth}
  \centering
  \includegraphics[width=.8\linewidth]{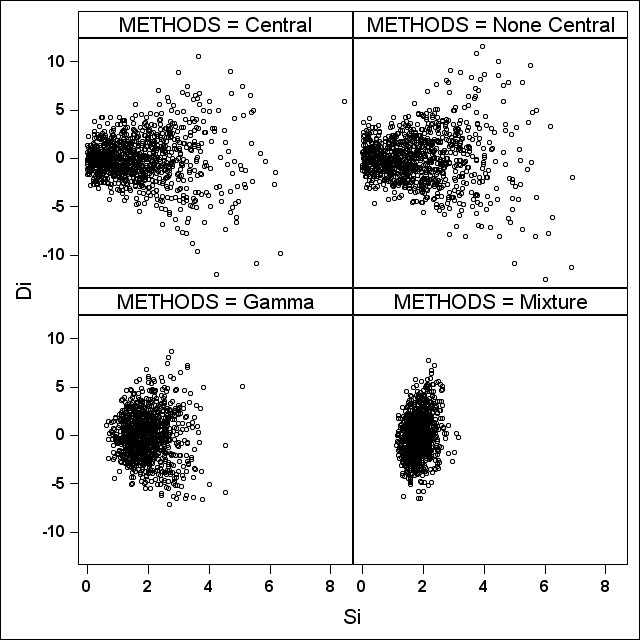}  
  \caption{$\tau_2^2=\log(2)$ and $\rho=0.7$}
  \label{fig:sub-third}
\end{subfigure}
\begin{subfigure}{.5\textwidth}
  \centering
  \includegraphics[width=.8\linewidth]{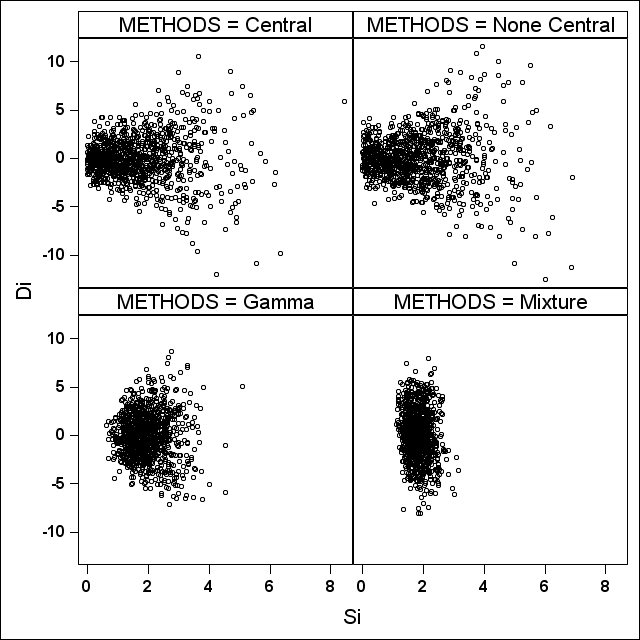}  
  \caption{$\tau_2^2=\log(2)$ and $\rho=-0.7$}
  \label{fig:sub-fourth}
\end{subfigure}
\caption{A scatter plot of $S_i$ against $D_i$ for four different simulation settings (i.e., Central chi-square, Gamma, None-central chi-square, and Mixture of chi-square) based on 100000 simulated studies.} 
\label{fig:fig}
\end{figure}

We will first discuss the simulation settings for our own aggregated
data meta-analysis. Then from these settings we will determine appropriate
settings for the simulation models from literature to make the comparison
of these simulations with our simulations as fair as possible (see Figure~\ref{fig:fig}). Then
after defining the settings, we will discuss the publication bias
approach and its settings that were used in the simulation study for
the publication bias methods. Our aim is to investigate the influence of different simulation models on the performance of various meta-analysis methods. We will study the pooled estimator of DerSimonian and Laird~\cite{DerSimonian1986} and the publication bias approaches PET-PEESE~\cite{Stanley2008,Stanley2014}, and Trim \& Fill~\cite{Duval2000, Duval2000a}. Details on these approaches can also be found in the Appendix.

\subsection{Simulation Settings}

We selected two levels for the number of studies in the simulated
meta-analysis ($m\in\{10,50\}$), but we chose to fix the parameters
for simulating the sample sizes within studies ($\lambda=100$, $a=b=1$,
$p=0.5$). This results in an expected number of sample sizes within
each study that is equal to $n_{i0}=n_{i1}=50$, but they could vary
strongly from study to study due to the overdispersed Poisson distribution
and vary within study due to a binomial division of the total sample
size $n_{i0}+n_{i1}$. We considered no treatment effect and two positive
levels of treatment effect ($\beta_{1}\in\{0,2,5\}$)\footnote{Note that we could exclude values for $\mu$ in model (\ref{eq:IPD model}),
since this parameter cancels out at an aggregated level when mean
differences are calculated.}. The residual variances for the two treatment groups were kept fixed
at the levels $\sigma_{0}^{2}=100$ and $\sigma_{1}^{2}=64$, respectively.
For the variance of treatment heterogeneity we selected three levels
$\tau^{2}\in\{0,2,5\}$.
The variance for the random heteroscedasticity was selected at two
levels, $\tau_{2}^{2}\in\{0,\log(2)\}$, to simulate effect sizes
with and without a random heteroscedasticity. Finally, we selected
three levels for the correlation coefficient between the heteroscedasticity
and heterogeneity $\rho\in\{-0.7,0,0.7\}$.
Clearly, when $\tau_{2}^{2}=0$, the correlation coefficient $\rho$
would not play any role in the simulation. Thus we will study $66=18+48$
different simulation settings (see Table~\ref{tab:sim_setting} for an overview of the simulation settings for one of the choices of $\beta_1$\footnote{The $I^2$'s for other choices of the $\beta_1$ will be very similar to what is being reported in Table~\ref{tab:sim_setting}.}). Each setting will simulate 1000 meta-analysis
studies.

\begin{table}[!htp]
\centering
\caption{An overview of the simulation settings and their corresponding mean $I^2$ estimated from the simulated data for $\beta_1=0$.}\label{tab:sim_setting}
\begin{adjustbox}{width=\textwidth}
\begin{tabular}{ccrrrrrrr}
  \toprule
  \multirow{3}[6]{*}{$m$} & \multirow{3}[6]{*}{$\tau^2$} & \multicolumn{1}{c}{\multirow{3}[6]{*}{$\tau_2^2$}} & \multicolumn{6}{c}{$I^2$ (5th percentile - 95th percentile)} \\
  \cmidrule{4-9}  &   &   & \multicolumn{1}{c}{\multirow{2}[4]{*}{Central}} & \multicolumn{1}{c}{\multirow{2}[4]{*}{Gamma}} & \multicolumn{1}{c}{\multirow{2}[4]{*}{Non-central}} & \multicolumn{3}{c}{Mixture} \\
  \cmidrule{7-9}  &   &   &   &   &   & \multicolumn{1}{c}{$\rho=-0.7$} & \multicolumn{1}{c}{$\rho=0$} & \multicolumn{1}{c}{$\rho=0.7$} \\
  \midrule
  \multirow{6}[12]{*}{10} & \multirow{2}[4]{*}{0} & \multicolumn{1}{r}{0} & 11.6 (0-47.6) & 11.5 (0-47.4) & 11.8 (0-46.8) & - & 11.0 (0-46.3) & - \\
  \cmidrule{3-9}  &   & $\log(2)$ & 11.6 (0-47.6) & 11.5 (0-47.4) & 11.8 (0-46.8) & - & 11.3 (0-48.1) & - \\
  \cmidrule{2-9}  & \multirow{2}[4]{*}{2} & \multicolumn{1}{r}{0} & 70.4 (10.0-98.2) & 36.0 (0-72.6) & 71.4 (12.8-98.5) & - & 31.6 (0-68.4) & - \\
  \cmidrule{3-9}  &   & $\log(2)$ & 64.3 (0-97.5) & 32.7 (0-70.6) & 65.3 (0-97.9) & 41.8 (0-74.9) & 31.5 (0-68.4) & 25.2 (0-63.5) \\
  \cmidrule{2-9}  & \multirow{2}[4]{*}{5} & \multicolumn{1}{r}{0} & 83.8 (51.4-99.3) & 57.2 (3.6-84.3) & 84.6 (52.2-84.6) & - & 51.2 (0-78.6) & - \\
  \cmidrule{3-9}  &   & $\log(2)$ & 79.4 (38.0-99.0) & 53.0 (0-82.0) & 80.2 (39.2-99.1) & 64.2 (21.4-85.7) & 52.2 (0-80.0) & 41.9 (0-74.8) \\
  \midrule
  \multirow{6}[12]{*}{50} & \multirow{2}[4]{*}{0} & \multicolumn{1}{r}{0} & 6.7 (0-26.5) & 6.7 (0-26.7) & 6.7 (0-26.2) & - & 7.0 (0-27.4) & - \\
  \cmidrule{3-9}  &   & $\log(2)$ & 6.7 (0-26.5) & 6.7 (0-26.7) & 6.7 (0-26.2) & - & 7.3 (0-27.2) & - \\
  \cmidrule{2-9}  & \multirow{2}[4]{*}{2} & \multicolumn{1}{r}{0} & 93.1 (78.2-99.7) & 44.1 (21.4-61.8) & 92.7 (77.9-99.7) & - & 36.0 (11.1-54.6) & - \\
  \cmidrule{3-9}  &   & $\log(2)$ & 90.7 (71.3-99.5) & 39.8 (15.5-58.5) & 90.4 (71.3-99.6) & 49.4 (29.3-64.7) & 37.1 (13.5-56.0) & 28.1 (0-49.8) \\
  \cmidrule{2-9} & \multirow{2}[4]{*}{5} & \multicolumn{1}{r}{0} & 97.0 (90.2-99.9) & 62.8 (47.0-75.6) & 96.9 (89.8-99.9) & - & 59.0 (43.2-70.9) & - \\
  \cmidrule{3-9}  &   & $\log(2)$ & 95.9 (86.6-99.8) & 66.8 (52.4-78.5) & 95.7 (86.0-99.8) & 71.2 (59.7-80.1) & 60.0 (45.3-72.6) & 49.7 (30.5-65.8) \\
  \bottomrule
  \end{tabular}%
  \end{adjustbox}
\end{table}

Relating our simulation settings to the parameters of model (\ref{eq:ADMA-random-effect}),
we obviously obtain that $\theta\in\{0,2,5\}$ and $\tau^{2}\in\{0,2,5\}$,
since they are direct translations of $\beta_{1}$ and $\tau^{2}$
given above. Thus we could draw a normally distributed random variable
$U_{i}$ having a mean equal to $\theta$ and a variance equal to
$\tau^{2}$ for all the simulation models from literature. Without
random heteroscedasticity ($\tau_{2}^{2}=0$), the residual variance
in (\ref{eq:ADMA-random-effect}) would become equal to $\sigma_{i}^{2}=\sigma_{0}^{2}/n_{i0}+\sigma_{1}^{2}/n_{i1}$.
Using the settings $\sigma_{0}^{2}=100$, $\sigma_{1}^{2}=64$, and
$n_{i0}=n_{i1}\approx50$, the variance $\sigma_{i}^{2}$ would become
on average equal to approximately $3.28$, but it will vary across
studies due to randomness in $n_{i0}$ and $n_{i1}$. Thus when instead
we will draw $\sigma_{i}^{2}$ directly from the central chi-square,
the non-central chi-square, or the gamma distribution, we need to
choose the distributional parameters such that the expected value
is close to $3.28$. For the central chi-square distribution we use
$\sigma_{i}^{2}=3.28\chi_{1}^{2}$, having a mean of $3.28$ and a
standard deviation of $3.28\sqrt{2}$. For the non-central chi-square
distribution we use $\sigma_{i}^{2}=3(0.3+Z_{i})^{2}$, with $Z_{i}\sim N(0,1)$.
The expected value is then equal to $3(1+0.09)=3.27$ and the standard
deviation is $3\sqrt{2(1+0.18)}$. For the gamma distribution we choose
the parameter settings $\Gamma(9,5)$ for standard error $S_{i}$,
leading to a mean of $9/5\approx\sqrt{3.28}$ and a standard deviation
of $3/5$.

In case of random heteroscedasticity ($\tau_{2}^{2}=\log(2)>0$),
the residual variance in (\ref{eq:Conditional_Di}) for the simulation
models from literature would just increase on average with a factor
$\sqrt{2}=\exp\{0.5\log(2)\}$ compared to the non-random heteroscedasticity
setting $(\tau_{2}^{2}=0$). These simulation models do not alter
the relation between $D_{i}$ and $S_{i}$ when a random heteroscedasticity
is introduced. Thus for a setting with random heteroscedasticity,
we should just multiply the random variables from the setting without
random heteroscedasticity with the factor $\sqrt{2}$ when we use
the simulation models with the chi-square distributions and $\sqrt[4]{2}$
when we use simulation model with the gamma distribution. Thus correlation
coefficient $\rho$, which affects the correlation between $D_{i}$
and $S_{i}$ in our simulation model, does not affect any of the simulation
models from literature.

\subsection{Publication Bias}

The performance of the Trim \& Fill and PET-PEESE method in literature
was investigated with the same data-driven publication bias selection
approach. All studies with a significant effect size are never being
excluded from the meta-analysis. For the remaining (non-significant)
studies, a standard uniform distributed random variable is generated
$U(0,1)$ for each study and the study is included in the meta-analysis
when the value of the uniform distributed random variable is less
than $\pi_{\mathrm{pub}}$.

The level of significance and the parameter $\pi_{\mathrm{pub}}$
are chosen such that the desired publication rate over all studies
is approximately 70\%. A study is considered significant when the
standardized effect size $D_{i}/S_{i}$ satisfies $D_{i}/S_{i}>z_{\alpha}$,
with $z_{q}$ the upper $q$\textsuperscript{th} quantile of the
standard normal distribution function. We used this $z$-test because
the aggregated data simulation models from literature did not implement
degrees of freedom $df_{i}$, ruling out the use of a $t$-test. In
this way a comparison of all aggregated data simulation models would
be fair. For $\theta=0$ and $\theta=2$ we used as significance level
of $\alpha=0.05$, but for $\theta=5$ we used the smaller significance
levels of $\alpha=0.0013$ (three sigma), since some settings had
more than 70\% significant studies. Thus publication bias depends
on the relative performance of other studies. We then pragmatically
tuned the parameter $\pi_{\mathrm{pub}}$ for the different settings
to obtain the target of 70\%. Note that $\pi_{\mathrm{pub}}$ would
also be different for the different simulation models, due to the
differences in generating the variance $S_{i}^{2}$.

\section{Results}
\subsection{DerSimonian and Laird}

The results of the four aggregated data simulation models for $m=10$
are provided in Table \ref{tab:DSL_m=00003D50_no}. Here we did not
include publication bias, since it is well-known that DerSimonian
and Laird's estimate would be biased~\cite{Duval2000a, Stanley2014}. The correlation coefficient $\rho$ for the
random heteroscedastic model was set to $\rho=0$ when we studied
heteroscedasticity.
\begin{table}[]
\caption{Comparison of aggregated data simulation models on bias ($\times 10^{-3}$) and coverage probability (in \%) for the DerSimonian and Laird random
effects estimator of the overall treatment effect ($m=10$; $\theta=2$; $\rho=0$)}
\label{tab:DSL_m=00003D50_no}
\begin{adjustbox}{width=\textwidth}
\begin{tabular}{llrrrrrrrr}
\toprule
\multirow{2}{*}{$\tau_2^2$} & \multirow{2}{*}{$\tau^2$} & \multicolumn{2}{c}{Central} & \multicolumn{2}{c}{Non-central} & \multicolumn{2}{c}{Gamma} & \multicolumn{2}{c}{Mixture} \\ \cmidrule(r){3-4}\cmidrule(r){5-6}\cmidrule(r){7-8}\cmidrule(r){9-10} 
 &  & Bias (MCSE) & 95\% CI & Bias (MCSE) & 95\% CI & Bias (MCSE) & 95\% CI & Bias (MCSE) & 95\% CI \\ \midrule
\multirow{3}{*}{0} & 0 & 8.5 (06.6) & 94.1 & 11.1 (06.9) & 94.6 & 26.3 (16.5) & 94.6 & 4.5 (18.3) & 94.6 \\
 & 2 & -7.0 (19.5) & 89.6 & 22.8 (19.7) & 90.7 & 12.7 (22.8) & 92.4 & 6.4 (23.1) & 94.7 \\
 & 5 & -14.9 (27.3) & 92.9 & 15.2 (27.4) & 93.1 & 9.2 (29.1) & 94.0 & 8.7 (28.8) & 95.1 \\ \midrule
\multirow{3}{*}{$\log(2)$} & 0 & 10.2 (07.9) & 94.1 & 13.3 (08.2) & 94.6 & 28.7 (18.0) & 94.6 & -6.1 (17.5) & 94.4 \\
 & 2 & -6.2 (20.7) & 89.1 & 28.7 (20.9) & 89.9 & 14.6 (24.0) & 92.2 & -7.9 (22.4) & 94.6 \\
 & 5 & -15.1 (28.4) & 91.9 & 22.2 (28.6) & 92.4 & 10.6 (30.1) & 93.7 & -11.1 (27.9) & 95.0 \\ \bottomrule
\end{tabular}
\end{adjustbox}
\end{table}

All simulation models show hardly any bias, but there is a small difference
in the coverage probability for the 95\% confidence intervals of the
pooled effect size. The simulation model with the mixture of chi-square
distributed variances shows a nominal coverage probability, while
the other simulation models underestimate the coverage probability
when there is heterogeneity. In case the number of studies increases,
the difference gets smaller, but remains present (data not shown).
The heteroscedasticity ($\rho=0$), which just increases the residual
variances for the simulation models (since there is no linear correlation
between $D_{i}$ and $S_{i}^{2}$), does not seem to affect any of
the results a lot.

\subsection{Trim \& Fill}

For the evaluation of the Trim \& Fill method, we implemented the
publication bias mechanism. For all aggregated data simulation models,
the average number of studies that were included was around 70\% (and
it ranged from 67\% to 73\%). Table \ref{tab:Trim=000026Fill-Homo}
shows the results on estimation bias and coverage probability for
the treatment effect of the Trim \& Fill approach under homoscedasticity
($\tau_{2}^{2}=0$) for $m=50$ studies. Results for $m=10$ are slightly
worse (data not shown).

\begin{table}[h]
\caption{Comparison of aggregated data simulation models on bias ($\times10^{-3})$
and coverage probability (in \%) for the Trim \& Fill estimator ($m=50$;
$\tau_{2}^{2}=0$).}
\label{tab:Trim=000026Fill-Homo}
\begin{adjustbox}{width=\textwidth}
\begin{tabular}{llrrrrrrrr}
\toprule
\multirow{2}{*}{$\theta$} & \multirow{2}{*}{$\tau^2$} & \multicolumn{2}{c}{Central} & \multicolumn{2}{c}{Non-central} &\multicolumn{2}{c}{Gamma} & \multicolumn{2}{c}{Mixture} \\ \cmidrule(r){3-4}\cmidrule(r){5-6}\cmidrule(r){7-8}\cmidrule(r){9-10} 
 &  & Bias (MCSE) & 95\% CI & Bias (MCSE) & 95\% CI & Bias (MCSE) & 95\% CI & Bias (MCSE) & 95\% CI \\ \midrule
\multirow{3}{*}{0} & 0 & 6.9 (02.0) & 97.6 & 8.9 (02.2) & 96.9 & 28.7 (10.4) & 90.2 & 89.4 (15.6) & 80.2 \\
 & 2 & 232.8 (13.0) & 74.8 & 178.9 (12.9) & 77.6 & 130.1 (17.6) & 83.3 & 173.8 (18.3) & 81.7 \\
 & 5 & 398.1 (19.6) & 70.3 & 314.9 (19.6) & 73.5 & 258.7 (22.7) & 80.3 & 288.5 (20.8) & 83.9 \\ \midrule
\multirow{3}{*}{2} & 0 & 10.0 (01.4) & 99.1 & 11.2 (01.5) & 97.7 & 192.8 (09.3) & 87.5 & 261.1 (14.2) & 79.0 \\
 & 2 & 261.2 (11.3) & 73.4 & 248.8 (11.2) & 76.4 & 308.3 (16.6) & 78.1 & 355.3 (16.6) & 78.5 \\
 & 5 & 495.3 (17.3) & 64.5 & 473.1 (17.6) & 64.0 & 434.5 (21.7) & 75.2 & 484.5 (19.9) & 78.2 \\ \midrule
\multirow{3}{*}{5} & 0 & 0.8 (01.3) & 98.8 & 1.2 (01.5) & 97.7 & 114.3 (08.8) & 90.8 & 391.9 (13.7) & 71.9 \\
 & 2 & -3.1 (11.9) & 79.7 & -20.3 (12.0) & 81.5 & 272.2 (15.4) & 77.6 & 509.9 (17.1) & 68.9 \\
 & 5 & 221.8 (17.8) & 76.7 & 198.1 (17.6) & 77.7 & 524.4 (20.6) & 69.3 & 675.8 (20.7) & 65.8 \\ \bottomrule
\end{tabular}
\end{adjustbox}
\end{table}

The simulation models show some different results with respect to
bias and coverage probability. The simulation models with the central
and non-central distribution do not show any real bias when heterogeneity
is absent, while the simulation model with the mixture of chi-square
distributions shows a small positive bias at treatment effects $\theta=2$
and $\theta=5$ (Trim \& Fill does not correct enough). The coverage
probability of the central and non-central chi-square are conservative,
but they are underestimated with the gamma and the mixture of chi-square
distributions. In the presence of heterogeneity, all simulation models
show real biases that increases with the size of heterogeneity. For
the central and non-central chi-square distribution, the absolute
bias is highest for treatment effect $\theta=2$, but for the mixture
of chi-square distributions the bias is highest at treatment effect
$\theta=5$. For the gamma distribution, the absolute bias is higher
at $\theta=2$ than at $\theta=5$ when heterogeneity is limited,
but smaller when heterogeneity is at the level of $\tau^{2}=5$. The
mixture of chi-square distributions generally provides the highest
bias with respect to the other simulation models.

Random heteroscedasticity with $\rho=0$ does not alter the results
observed in Table \ref{tab:Trim=000026Fill-Homo}, since it merely
increases the residual variance for the simulation models. However,
when we select $\rho=-0.7$ or $\rho=0.7$, the correlation between
heterogeneity and heteroscedasticity seems to change the results for
our aggregated data simulation model (see Table \ref{tab:Trim=000026Fill-Hetero}).
A negative correlation increases the bias and lowers the coverage,
while a positive correlation eliminates the bias or changes it to
a negative bias, but always with highly liberal coverage probabilities.
A negative correlation $\rho=-0.7$ introduces a negative correlation
between $D_{i}$ and $S_{i}$ and diminishes the positive correlation
between $D_{i}$ and $S_{i}$ that was induced by the publication
bias mechanism, masking an asymmetry in the funnel plot that was introduced
by the publication bias. A positive correlation $\rho=0.7$ enhances
the publication bias, making the Trim \& Fill approach correct stronger.
For $\theta=0$ the positive bias in Table \ref{tab:Trim=000026Fill-Homo}
under homoscedasticity (or under $\rho=0$) is turned into a negative
bias, while the observed positive bias for $\theta\in\{2,5\}$ in
Table \ref{tab:Trim=000026Fill-Homo} is almost eliminated (Table
\ref{tab:Trim=000026Fill-Hetero}).

\begin{table}[h]
\caption{Bias ($\times10^{-3})$ and coverage probability (in \%) for the Trim
\& Fill estimator under the mixture of chi-square distributions with heteroscedasticity ($m=50$).}
\label{tab:Trim=000026Fill-Hetero}
\centering
  \begin{tabular}{llrrrrrr}
  \toprule
  \multirow{2}{*}{$\theta$} & \multirow{2}{*}{$\tau^2$} & \multicolumn{2}{c}{$\rho=-0.7$} & \multicolumn{2}{c}{$\rho=0$} & \multicolumn{2}{c}{$\rho=0.7$} \\ \cmidrule(r){3-4} \cmidrule(r){5-6} \cmidrule(r){7-8} 
   &  & Bias (MCSE) & 95\% CI & Bias (MCSE) & 95\% CI & Bias (MCSE) & 95\% CI \\ \midrule
  \multirow{2}{*}{0} & 2 & 609.7 (19.9) & 69.0 & 172.4 (18.6) & 81.5 & -319.3 (17.7) & 74.7 \\
   & 5 & 796.6 (24.4) & 70.9 & 285.9 (21.3) & 82.5 & -344.9 (20.9) & 77.9 \\ \midrule
  \multirow{2}{*}{2} & 2 & 755.0 (18.4) & 57.5 & 334.4 (17.0) & 78.7 & -121.3 (16.2) & 85.8 \\
   & 5 & 942.0 (22.7) & 60.0 & 457.9 (20.2) & 77.9 & -123.7 (19.0) & 87.3 \\ \midrule
  \multirow{2}{*}{5} & 2 & 873.9 (17.9) & 46.3 & 437.7 (16.5) & 70.1 & 5.7 (14.9) & 87.3 \\
   & 5 & 1126.6 (22.8) & 48.0 & 587.1 (20.2) & 67.6 & 18.3 (18.0) & 86.9 \\ \bottomrule
  \end{tabular}
\end{table}

\subsection{PET-PEESE}

For the evaluation of the PET-PEESE estimator, we also implemented
the publication bias mechanism. Since we used the same data as for
the Trim \& Fill approach, the average number of studies that were
included was around 70\% (and it ranged from 67\% to 73\%). Table
\ref{tab:PETPEESE-Homo} shows the results on estimation bias and
coverage probability for the treatment effect of the PET-PEESE under
homoscedasticity ($\tau_{2}^{2}=0$) for $m=50$ studies. Results
for $m=10$ are less severe for the simulation models with the central
and non-central chi-square distribution when the treatment effect
is restricted ($\theta\in\{0,2\}$), but in all other cases the results
are more extreme for $m=10$ than for $m=50$. 

\begin{table}[h]
\caption{Comparison of aggregated data simulation models on bias ($\times10^{-3})$
and coverage probability (in \%) for the PET-PEESE estimator ($m=50$;
$\tau_{2}^{2}=0$).}
\label{tab:PETPEESE-Homo}
\begin{adjustbox}{width=\textwidth}
\begin{tabular}{llrrrrrrrr}
\toprule
\multirow{2}{*}{$\theta$} & \multirow{2}{*}{$\tau^2$} & \multicolumn{2}{c}{Central} & \multicolumn{2}{c}{Non-central} &\multicolumn{2}{c}{Gamma} & \multicolumn{2}{c}{Mixture} \\ \cmidrule(r){3-4}\cmidrule(r){5-6}\cmidrule(r){7-8}\cmidrule(r){9-10} 
 &  & Bias (MCSE) & 95\% CI & Bias (MCSE) & 95\% CI & Bias (MCSE) & 95\% CI & Bias (MCSE) & 95\% CI \\ \midrule
\multirow{3}{*}{0} & 0 & 1.2 (02.0) & 98.1 & 1.9 (02.2) & 97.1 & -20.9 (23.4) & 97.2 & 106.3 (102.7) & 95.7 \\
 & 2 & 212.3 (33.0) & 26.8 & 241.2 (31.4) & 26.7 & 78.7 (35.8) & 91.0 & 192.2 (129.6) & 95.6 \\
 & 5 & 352.6 (51.6) & 25.7 & 400.4 (49.6) & 25.9 & 208.6 (48.9) & 88.3 & 337.5 (162.0) & 95.6 \\ \midrule
\multirow{3}{*}{2} & 0 & 2.2 (01.1) & 97.7 & 2.5 (01.3) & 95.4 & 14.0 (19.2) & 93.7 & 129.4 (097.9) & 94.9 \\
 & 2 & 130.7 (31.2) & 26.3 & 157.6 (29.6) & 26.4 & 39.5 (32.1) & 86.1 & 199.6 (124.9) & 94.5 \\
 & 5 & 382.1 (48.5) & 24.6 & 423.4 (46.5) & 23.5 & 127.2 (44.8) & 81.1 & 273.8 (156.1) & 94.5 \\ \midrule
\multirow{3}{*}{5} & 0 & -5.0 (01.1) & 96.2 & -5.4 (01.4) & 94.3 & -41.5 (12.5) & 94.6 & -281.5 (088.1) & 94.7 \\
 & 2 & -25.4 (33.2) & 26.8 & 16.1 (30.4) & 28.1 & 62.1 (24.4) & 87.7 & -228.2 (111.5) & 95.5 \\
 & 5 & 37.0 (50.9) & 27.1 & 114.1 (46.3) & 27.1 & 177.8 (40.5) & 82.4 & -113.2 (141.7) & 95.0 \\ \bottomrule
\end{tabular}
\end{adjustbox}
\end{table}

For the PET-PEESE we see biases for the simulation models with the
central and non-central chi-square distributions similar to their
biases of the Trim \& Fill approach, but the coverage probabilities
are much smaller when heterogeneity is present. The central and non-central
chi-square simulation models simulate studies with very small standard
errors (i.e., mimicking extremely large studies in a meta-analysis).
These extreme studies, in combination with an independent heterogeneous
study effect, have a strong influence on the estimation of the effect
size and its standard error in a single meta-analysis. Since the heterogeneity
is independent of the standard error, the bias is less affected when
averaged out over many meta-analyses, but the confidence interval
of the pooled study effect size from a single meta-analysis will not
capture the true parameter. The PET-PEESE does not incorporate the
heterogeneity in the weights and therefore underestimate the standard
error, but the Trim \& Fill does account somewhat for the heterogeneity.

The simulation model with the gamma distribution and the mixture of
chi-square distributions seem to show less bias with the PET-PEESE
than with the Trim \& Fill. The bias with PET-PEESE is limited and
acceptable for the simulation model with the gamma distribution, although
the coverage probability is liberal when heterogeneity is present.
The coverage probability for the simulation model with a mixture of
chi-square distributions are very close to nominal, even though a
small bias is frequently present. The change from a positive bias
for treatment effects $\theta\in\{0,2\}$ to a negative bias when
$\theta=5$, is difficult to explain, but could be related to the
high number of significant effect sizes. For $\theta=5$, Egger's
part hardly plays a role in estimation of the pooled treatment effect,
which is not the case for smaller treatment effects. Also the other
simulation models show a decline in the bias when the treatment $\theta$
changes from $\{0,2\}$ to $5$, suggesting it is related to the PET-PEESE
approach. 

\begin{table}[h]
\caption{Bias ($\times10^{-3})$ and coverage probability (in \%) for the PET-PEESE
estimator under the mixture of chi-square distributions with heteroscedasticity ($m=50$).}
\label{tab:PETPEESEl-Hetero}
\centering
\begin{tabular}{llrrrrrr}
  \toprule
  \multirow{2}{*}{$\theta$} & \multirow{2}{*}{$\tau^2$} & \multicolumn{2}{c}{$\rho=-0.7$} & \multicolumn{2}{c}{$\rho=0$} & \multicolumn{2}{c}{$\rho=0.7$} \\ \cmidrule(r){3-4} \cmidrule(r){5-6} \cmidrule(r){7-8} 
   &  & Bias (MCSE) & 95\% CI & Bias (MCSE) & 95\% CI & Bias (MCSE) & 95\% CI \\ \midrule
   \multirow{2}{*}{0} & 2 & 2560.5 (55.4) & 76.6 & 238.3 (66.5) & 95.5 & -2112.5 (49.3) & 78.5 \\
   & 5 & 3909.2 (66.4) & 66.4 & 364.0 (83.5) & 94.9 & -3057.6 (48.8) & 64.2 \\ \midrule
  \multirow{2}{*}{2} & 2 & 2140.9 (45.2) & 66.9 & 153.9 (59.7) & 93.7 & -2487.0 (60.5) & 78.0 \\
   & 5 & 3292.8 (56.9) & 56.2 & 279.6 (77.9) & 93.5 & -3674.5 (65.9) & 63.4 \\ \midrule
  \multirow{2}{*}{5} & 2 & 1680.8 (46.4) & 69.0 & -118.8 (56.3) & 93.8 & -2530.4 (66.0) & 76.2 \\
   & 5 & 2729.8 (58.7) & 58.1 & -102.4 (71.0) & 93.0 & -3905.5 (76.9) & 64.1 \\ \bottomrule
  \end{tabular}
\end{table}

Similar to the Trim \& Fill method, a random heteroscedasticity with
$\rho=0$ does not alter the results observed in Table \ref{tab:PETPEESE-Homo}
a lot, since it does not change the correlation between $D_{i}$ and
$S_{i}^{2}$ compared to the setting with homoscedasticity. However,
when $\rho=-0.7$ or $\rho=0.7$, we see a similar pattern as for
the Trim \& Fill approach, but now the heteroscedasticity has a stronger
effect (see Table \ref{tab:PETPEESEl-Hetero}). A negative correlation
$\rho=-0.7$ reduces the positive correlation between $D_{i}$ and
$S_{i}$ induced by the publication bias mechanism, failing the PET-PEESE
to correct for publication bias. A positive correlation $\rho=0.7$
enhances the publication bias and making the PET-PEESE over correct.

\section{Summary and Discussion}

We discussed four simulation models for generating aggregated data
in a meta-analysis study. All four models use the well-known random
effects model~\cite{DerSimonian1986} to generate study effect
sizes, but each model used their own distribution function for generating
the standard error of the study effect size (a central chi-square,
a non-central chi-square, a gamma, and a mixture of chi-square distributions).
We showed that the mixture of chi-square distributions would follow
naturally from a mixed model for individual participant data (IPD),
but the other three distributions could not be formulated directly
from such an IPD model. The simulation models with a central chi-square,
non-central chi-square, and gamma distribution indirectly created
a dependency between the study effect size and its standard error,
but without introducing a linear dependency. For the mixture of chi-square
distributions, a dependency between the study effect size and its
standard error was created through a random heteroscedasticity, in
line with research on hierarchical heteroscedastic multi-level models,
which gave a zero correlation only when the heteroscedasticity is
unrelated to the study heterogeneity. The simulation models with a
central and non-central chi-square distribution, simulated studies
with very small and relatively large standard errors, mimicking meta-analyses
with a large variety of study sizes, while the gamma distribution
and the mixture of chi-square distributions typically showed standard
errors that belong to a smaller variety in study sizes.

Our simulation study showed that the choice of simulation model affects
the conclusion of the applied meta-analysis method. The well-known
liberal coverage probabilities of the DerSimonian and Laird method
of pooled effect size~\cite{Veroniki2015, Brockwell2001, Hardy1996} was observed with the simulation
models using the central and non-central chi-square distribution,
and to a lesser extent with the gamma distribution and the mixture of chi-square distributions. The simulation model
using either the gamma distribution or the mixture of chi-square distributions would then lead to the conclusion that the DerSimonian and Laird method had close-to-nominal coverage probabilities.

All simulation models showed some biases for the PET-PEESE approach
when heterogeneity is present and the treatment effects are moderate
to none. This influence of heterogeneity on the performance of PET-PEESE
is in line with literature~\cite{Stanley2014, Alinaghi2018}. However, they applied a different simulation model,
where the regression equations (\ref{eq:EGGER TEST}) and (\ref{eq:Moreno})
are directly simulated using uniform and normal distributions. Their
biases are therefore somewhat different from ours, but this strengthen
our point that performances of meta-analysis methods are sensitive
to simulation models. Indeed, in our own simulation study for the
PET-PEESE we demonstrated that the simulation model with the gamma
distribution showed the smallest bias and provided acceptable biases
across all settings. Contrary, the simulation model with the mixture
of chi-square distributions gave biases of the pooled effect size
even when heterogeneity was not present, which was not seen with the
other three simulation models. Finally, PET-PEESE was sensitive to
standard errors that belong to a larger variety of study sizes, since
it gave very small coverage probabilities for the pooled effect size
when the central and non-central chi-square distributions were used
with study heterogeneity. The simulated meta-analyses typically contained
extremely small standard errors that caused an underestimation of
the variance of the pooled effect size. To our knowledge, this sensitivity
to study sizes has not been presented in literature before.

For the Trim \& Fill approach, the simulation model with the gamma
distribution showed larger biases in the pooled treatment effect than
for the PET-PEESE approach. This is in line with findings that PET-PEESE
seems to outperform the Trim \& Fill approach~\cite{Stanley2014}. However, the simulation models with the central and non-central
chi-square distributions did not show any large discrepancies between
Trim \& Fill and PET-PEESE, arguing that Trim \& Fill may not necessarily
be worse than PET-PEESE. On the other hand, the simulation models
with the mixture of chi-square distributions showed substantial positive
biases with the Trim \& Fill approach and moderate negative biases
with the PET-PEESE approach when the treatment effect is strong, an
observation not mentioned earlier. Interestingly though, this simulation
model with a mixture of chi-square distributions always showed coverage
probabilities very close to nominal for all settings for both the
Trim \& Fill and the PET-PEESE approach, despite the observed biases,
contrary to the other simulation models which showed (very) liberal
coverage probabilities when heterogeneity is present. Indirectly,
the variability in coverage probabilities also showed that the simulation
model affects the mean squared error, since the coverage probability
for the PET-PEESE is much smaller than for the Trim \& Fill approach
when the simulation models with central and non-central chi-square
distributions are used with heterogeneous effect sizes. Thus mean
squared errors are not just affected by publication bias methods~\cite{Stanley2014}, but also by the choice of simulation model.

When we introduce random heteroscedasticity, an element not studied
in literature before, the Trim \& Fill and PET-PEESE approach may
fail completely in their estimation of the pooled treatment effect.
This effect could only be observed with our simulation model, since
the simulation models with the central chi-square, non-central chi-square,
and the gamma distribution do not have a mechanism to change the joint
distribution of the study effect and its standard error. The failure
of the publication bias method is not unexpected, since the linear
correlation between the study effect and its standard error is influenced
by the heteroscedasticity, which is unrelated to publication biases,
and confuses the methods for publication bias.

One of the limitations of our study is that we have only considered one study effect (mean difference) for aggregated data meta-analysis that is calculated from one specific heteroscedastic IPD model. Our selected IPD model can also be used to generate binary outcomes by either using a threshold on the continuous response or by using a link function that would change the continuous response into a binary response, leading to several 2x2 contingency tables~\cite{almalik2018testing}. The cell counts can then be used to create an alternative study effect $D_i$ (e.g., odds ratio) with its appropriate standard error $S_i$, but their (joint) distribution would be currently unknown and may be work for future research. Alternatively, cell counts in 2x2 contingency tables can also be generated differently~\cite{sidik_comparison_2007,berkey_random-effects_1995,platt_generalized_1999,knapp_improved_2003} using a study-specific effect size at the aggregated level according to the random effects model~\eqref{eq:Effect_size}: and with an event probability of the treatment arm based on a logit model. This alternative IPD simulation approach showed that DerSimonian and Laird approach could lead to a biased estimate of the between study variability. Thus the choice of study effect and the different IPD models lead to different ways of simulating aggregated data for studying meta-analysis approaches and consequently could results in a variety of distributions for $(D_i, S_i^2, df_i)$ that may deviate from the choices we studied in this paper (see also the discussion of Jackson and White~\cite{jackson_when_2018} on the hidden distributional assumptions for the aggregated statistics used in meta-analyses). It emphasizes the importance of the choice of simulation model for generating aggregated measures of effect.

Our restricted investigation of simulation models already demonstrated that the choice of simulation model for aggregated
data meta-analyses can have an influence on the conclusion of how
well a particular meta-analysis approach performs. Most publications
in literature do not give any or strong arguments for their choice
of simulation models and could therefore implicitly bias their results
or conclusions. We recommend the use of multiple simulation models
when meta-analysis approaches are being studied to provide a more
fair view of the performance of the meta-analysis approach. Otherwise
we recommend that researchers provide a strong argument for their
choice of simulation model for the aggregated data and show that it
would make sense in reality, since not all simulation models we studied
had a clear interpretation to statistical models at IPD.

\section*{Appendix: Meta-Analysis Methods}

In this section we briefly describe the three meta-analysis approaches
that we study in our simulation: the DerSimonian and Laird pooled
analysis approach and two publication bias adjustment methods: Trim
\& Fill and PET-PEESE.

\subsection*{DerSimonian and Laird Method}

DerSimonian and Laird~\cite{DerSimonian1986} assumed that the study effect size $D_{i}$
follows model (\ref{eq:ADMA-random-effect}) in combination with the
normality assumptions on the random effects and with a known variance
$\sigma_{i}^{2}$ equal to the observed $S_{i}^{2}$. The pooled estimate
for $\theta$ is given by the weighted average $\hat{\theta}_{DSL}=\sum_{i=1}^{m}w_{i}D_{i}/\sum_{i=1}^{m}w_{i}$,
with weight $w_{i}$ equal to $w_{i}=[\hat{\tau}^{2}+S_{i}^{2}]^{-1}$
and $\hat{\tau}^{2}$ and estimator of the variance component for
the study effect heterogeneity. They proposed the estimator given
by
\[
\hat{\tau}_{DSL}^{2}=\max\left\{ 0,\frac{Q-\left(m-1\right)}{\sum_{i=1}^{m}S_{i}^{-2}-\sum_{i=1}^{m}S_{i}^{-4}/\sum_{i=1}^{m}S_{i}^{-2}}\right\} ,
\]

with Cochran's $Q$-statistic given by $Q=\sum_{i=1}^{m}[S_{i}^{-2}(D_{i}-\bar{D})^{2}]$
and with $\bar{D}$ the weighted average given by $\bar{D}=\sum_{i=1}^{m}[S_{i}^{-2}D_{i}]/\sum_{i=1}^{m}S_{i}^{-2}$.
The accompanied standard error $S$ of the pooled estimate $\hat{\theta}_{DSL}$
is given by $S^{2}=1/\sum_{i=1}^{m}w_{i}$. DerSimonian and Laird~\cite{DerSimonian1986} are not very clear on how to calculate confidence intervals,
but based on the work of Cochran~\cite{Cochran1954}, we assume that the degrees
of freedom of $S$ is equal to $m-1$ and use
\[
\hat{\theta}_{DSL}\pm t_{\alpha/2,m-1}^{-1}S
\]
for the $(1-\alpha)100\%$ confidence interval, with $t_{q,d}^{-1}$
the $(1-q)$ upper quantile of the $t$-distribution with $d$ degrees
of freedom. To obtain DerSimonian and Laird's pooled estimates $\hat{\theta}_{DSL}$
with its 95\% confidence limits we applied the R package ``meta''~\cite{Schwarzer2007}.

\subsection*{Trim \& Fill Method}

The Trim \& fill method has been described in detail in Duval and
Tweedie~\cite{Duval2000, Duval2000a}. In short, studies are first ranked based on
their distance from the pooled treatment effect estimated by the random
effects model (\ref{eq:ADMA-random-effect}), i.e., ranking distances
$|D_{i}-\hat{\theta}|$. Next, the number of unobserved studies is
estimated using for instance estimator $L_{0}=[4T_{m}-m(m+1)]/[2m-1]$,
where $T_{m}$ is the Wilcoxon rank sum test statistic estimated from
the ranks of studies with $D_{i}>\hat{\theta}$ (here we assume that
$\hat{\theta}$ is positive and it is more likely that studies with
effect sizes below $\hat{\theta}$ are potentially missing). We then
trim off the $L_{0}$ most extreme studies (i.e., studies with positive
effect sizes furthest away from zero) and re-estimate the pooled treatment
effect $\hat{\theta}$ without these studies. Then \textbf{all} studies
are ranked again, based on their distance to the new pooled estimate,
and $L_{0}$ is recomputed. This procedure is repeated until it stabilizes
($L_{0}$ does not change anymore) and we obtain a final estimate
$\hat{\theta}$ and a final estimate $L_{0}$ of the number of studies
missing. Then we impute $L_{0}$ studies by mirroring the $L_{0}$
studies with the highest effect sizes around the final estimate $\hat{\theta}$
and provide it with the standard error $S_{i}$ from the mirrored
study. After imputation, a final pooled estimate with standard error
is provided using the random effects model on all $m+L_{0}$ studies.
We used the function ``trimfill'' in the R package ``metafor'',
with the average treatment effect estimated using the random effects
model~\cite{Viechtbauer2010}.

\subsection*{PET-PEESE Method}

The PET-PEESE method is based on the understanding of the bias function
in case publication bias is present. Under certain conditions, this
bias function can be determined explicitly~\cite{Stanley2014}, but it is a complicated function of the true effect size and
approximations are needed. One approximation is based on Egger test~\cite{Egger1997}, which investigates the linear relation
between the $t$-value $T_{i}=D_{i}/S_{i}$ and the precision $S_{i}^{-1}$,
i.e.,
\begin{equation}
T_{i}=\alpha_{0}+\alpha_{1}S_{i}^{-1}+e_{i},\label{eq:EGGER TEST}
\end{equation}
with $\alpha_{0}$ and $\alpha_{1}$ the intercept and slope parameter,
respectively, and with residual $e_{i}\sim N\left(0,\nu^{2}\right)$.
An intercept deviating from zero ($\alpha_{0}\neq0$) would indicate
publication bias, while the slope in (\ref{eq:EGGER TEST}) represents
the effect size (i.e., $\alpha_{1}=\theta$). However, Stanley and
Doucouliagos~\cite{Stanley2014} demonstrated that the OLS estimator $\hat{\alpha}_{1}$
from (\ref{eq:EGGER TEST}) would be a biased estimator when it deviates
from zero. Thus when null hypothesis $H_{0}:\alpha_{1}=0$ is rejected,
they proposed to use the weighted linear regression between study
effect $D_{i}$ and variance $S_{i}^{2}$, i.e.,
\begin{equation}
D_{i}=\gamma_{0}+\gamma_{1}S_{i}^{2}+e_{i},\label{eq:Moreno}
\end{equation}
with weights $S_{i}^{-2}$ and report the estimate $\hat{\gamma}_{0}$
as the overall treatment effect $\theta$. This part is called the
Precision Effect Estimate with SE (PEESE). In case null hypothesis
$H_{0}:\alpha_{1}=0$ is not rejected, the reported estimate is $\hat{\alpha}_{1}$,
which is referred to as the Precision Effect Test (PET). Note that
model (\ref{eq:Moreno}) was suggested earlier by Moreno \textit{et
al.}~\cite{Moreno2011} We used Procedure GLM in SAS to carry out the PET-PEESE
method.

\begin{acks}
  We thank the two anonymous reviewers whose comments/suggestions helped improve and clarify this manuscript.
\end{acks}

\begin{dci}
  The Authors declare that there is no conflict of interest
\end{dci}

\begin{funding}
  The authors ERvdH and OA disclosed receipt of the following financial support for the research, authorship and/or publication of this article: This work was supported by the Netherlands Organization for Scientific Research [grant number 023.005.087]. The author ZZ received no financial support for the research, authorship, and/or publication of this article.
\end{funding}

\end{document}